in
\pdfoutput=1
\documentclass[%
prx,twocolumn,
notitlepage,
superscriptaddress,
 amsmath,amssymb,
 aps,
floatfix,
longbibliography
]{revtex4-2}

\usepackage[T1]{fontenc}
\usepackage[latin9]{inputenc}
\usepackage{babel}

\usepackage{amsmath,amssymb,graphicx}
\usepackage[capitalize]{cleveref}
\usepackage{bm}
\usepackage{gensymb}

\usepackage{enumerate}% http://ctan.org/pkg/enumerate
\usepackage{geometry}
\usepackage[]{mdframed} %framing text \begin{mdframed}
\usepackage[normalem]{ulem} %Strikethrough text with \sout
\usepackage{import} %Different tex files import

\definecolor{mvhPink}{rgb}{0.7,0.0,0.7}
\definecolor{dkGreen}{rgb}{0.0,0.7,0.}
\definecolor{ccBlue}{rgb}{0.0,0.67,1}
\definecolor{mdOrange}{rgb}{0.8,0.4,0}

\begin{document}

%\thispagestyle{plain}

%    \title{Nonlocal metasurfaces with giant tunability enabled by kirigami}
        \title{Mechano-optical metasurfaces}
\author{Freek van Gorp}
\affiliation{Institute of Physics, Universiteit van Amsterdam, 1098 XH Amsterdam, The Netherlands}
\author{Wenfeng Liu}
\affiliation{Institute of Physics, Universiteit van Amsterdam, 1098 XH Amsterdam, The Netherlands}
\author{Corentin Coulais}
\affiliation{Institute of Physics, Universiteit van Amsterdam, 1098 XH Amsterdam, The Netherlands}
\author{Jorik van de Groep}
\affiliation{Institute of Physics, Universiteit van Amsterdam, 1098 XH Amsterdam, The Netherlands}

\begin{abstract}
Tunable metasurfaces enable active and on-demand control over optical wavefronts through reconfigurable scattering of resonant nanostructures. Here, we present novel insights inspired by mechanical metamaterials to achieve giant tunability in mechano-optical metasurfaces where the mechanical metamaterial and optical metasurfaces are integrated in a single nanopatterned material. In a first design, judiciously engineered cuts in a flexible substrate enable large, strain-induced extension of the inter-particle spacing, tuning a high quality-factor resonance in a silicon nanoparticle array across a very broad spectral range. In a second design, we eliminate the substrate and demonstrate a nanopatterned silicon membrane that simultaneously functions as a mechanical metamaterial and an optical metasurface with large tunability. Our results highlight a promising route toward active metasurfaces, with potential applications in tunable filters, reconfigurable lenses, and dynamic wavefront shaping.
\end{abstract}
\maketitle

\emph{Introduction. ---} Molding the flow of light and controlling mechanical motion are daunting challenges for science and technology.
Metamaterials have recently emerged as exciting platforms for doing so 
in unprecedented ways.
On the one hand, photonic metamaterials employ suitable arrangements of interacting optical resonators to steer~\cite{Berini_2022}, filter~\cite{Wang_Wen_Deng_Li_Yang_2023}, and shape optical wavefronts~\cite{Chen_Taylor_Yu_2016b,kamali2018review}. On the other hand, mechanical metamaterials use the careful geometrical design of compliant hinges and rotating elements~\cite{Bertoldi_Vitelli_Christensen_van_Hecke_2017,shapemorphingreview2025} to combine low density, strength and energy dissipation~\cite{schaedler2011ultralight,Liu_Janbaz_Dykstra_Ennis_Coulais_2024}, 
to compute~\cite{Yasuda_Buskohl_Gillman_Murphey_Stepney_Vaia_Raney_2021, Bense_van_Hecke_2021a}, and to exhibit advanced shape morphing capabilities~\cite{shapemorphingreview2025,Coulais_Teomy_de_Reus_Shokef_van_Hecke_2016,Coulais_Sabbadini_Vink_van_Hecke_2018,Choi_Mahadevan_NatMat2019,Gladman_Matsumoto_Mahadevan_Lewis_2017,Gao_Bico_Roman_2023,Meeussen_van_Hecke_2023,Melancon_Gorissen_Bertoldi_2021,Smart_Pearson_Liang_Lim_Abdelrahman_Monticone_Cohen_McEuen_2024}. These two domains manipulate fundamentally different physical fields, operate across vastly different time and length scales and use markedly different mechanisms---resonances versus internal rotations. Yet in both, one confronts the same grand challenge of harnessing the collective response of their constitutive building blocks. From this perspective, a natural question arises: is it possible to merge these two fields? Can the exceptional tunability of mechanical metamaterials be harnessed to enhance our ability to manipulate light?

This prospect is particularly promising as the mechanical manipulation of optical resonators enables significant tunability of their optical response, and opens new possibilities for dynamic light control. Conventionally, such mechanical tuning of metasurfaces is achieved through relatively simple methods, such as stretching flexible substrates~\cite{Ee_Agarwal_2016,Pryce_Aydin_Kelaita_Briggs_Atwater_2010,Tseng_2017_mech_tune_full_visible,Zhang_Jing_Wu_Fan_Yang_Wang_Song_Xiao_2019,Shen_2014,GUAN2025416867} or employing out-of-plane deformations via MEMS technology~\cite{Zheludev_Plum_2016,holsteen2019}. However, mechanical metamaterials offer a distinct advantage by providing unique degrees of freedom through the internal rotations of their constituent building blocks, enabling more intricate and precise control over the system's geometry~\cite{Liu_Du_Li_Lu_Li_Fang_2018}. While initial demonstrations of light manipulation with metamaterial platforms have shown great promise~\cite{nano15010061,Xu_Wang_Kim_Shyu_Lyu_Kotov_2016b,Zheng_Chen_Yang_Wu_Qu_Zhao_Jiang_Feng_2021,Phon_Jeong_Lim_2022}, they have been restricted to macroscale unit cells and operational frequencies in the GHz-THz regime, rely on resonant particles supported by mechanically limited substrates, and employ localized resonances only.

Here, we address these limitations by combining the unique sensitivity of nonlocal metasurfaces~\cite{Lawrence_Barton_Dixon_Song_van_de_Groep_Brongersma_Dionne_2020,Song_van_Groep_Kim_Brongersma_2021,Overvig_Alu_2022} to inter-particle spacing and orientation with the large internal rotations offered by flexible mechanical metamaterials~\cite{shapemorphingreview2025, Zhai_Wu_Jiang_2021, Bertoldi_Vitelli_Christensen_van_Hecke_2017, Coulais_NatPhys2018, Lamoureux_NatComm2015, Rafsanjani_Bertoldi_2017, Choi_Mahadevan_NatMat2019} to demonstrate a nanopatterned membrane that simultaneously functions as an optical metasurface and a mechanical metamaterial --- a mechano-optical metasurface. 
We exploit these large internal rotations to change the distances and angles between resonant nanoparticles on-the-fly, which in turn allows us to dynamically tune the resonance condition and associated optical filtering properties of the optical metasurface, and to showcase the synergy between mechanical and optical functionalities in a single platform. Our approach augments the toolbox of reconfigurable metasurfaces~\cite{Pryce_Aydin_Kelaita_Briggs_Atwater_2010, Tseng_2017_mech_tune_full_visible, Zhang_Jing_Wu_Fan_Yang_Wang_Song_Xiao_2019, Shen_2014, Xu_Wang_Kim_Shyu_Lyu_Kotov_2016b, Howes_Zhu_Curie_Avila_Wheeler_Haglund_Valentine_2020, zhang_electrically_2021, cotrufo_reconfigurable_2024, Li_van_de_Groep_Talin_Brongersma_2019, Huang_Lee_Sokhoyan_Pala_Thyagarajan_Han_Tsai_Atwater_2016, Siegel_Kim_Fortman_Wan_Kats_Hon_Sweatlock_Jang_Brar_2024, Park_Kang_Kim_Liu_Brongersma_2016, Kaissner_Li_Lu_Li_Neubrech_Wang_Liu_2021, lewi_ultrawide_2017, wang_electrical_2021, Liu_Du_Li_Lu_Li_Fang_2018, Phon_Jeong_Lim_2022, Zheng_Chen_Yang_Wu_Qu_Zhao_Jiang_Feng_2021, Ee_Agarwal_2016,Chen_Ai_Wong_2020a} and opens avenues for the synergistic design of mechanical and optical functionalities. 

\begin{figure*}[t!]
%\hspace{-2.2cm}
\centering
\includegraphics[width=2\columnwidth,trim=0cm 5cm 0cm 0cm]{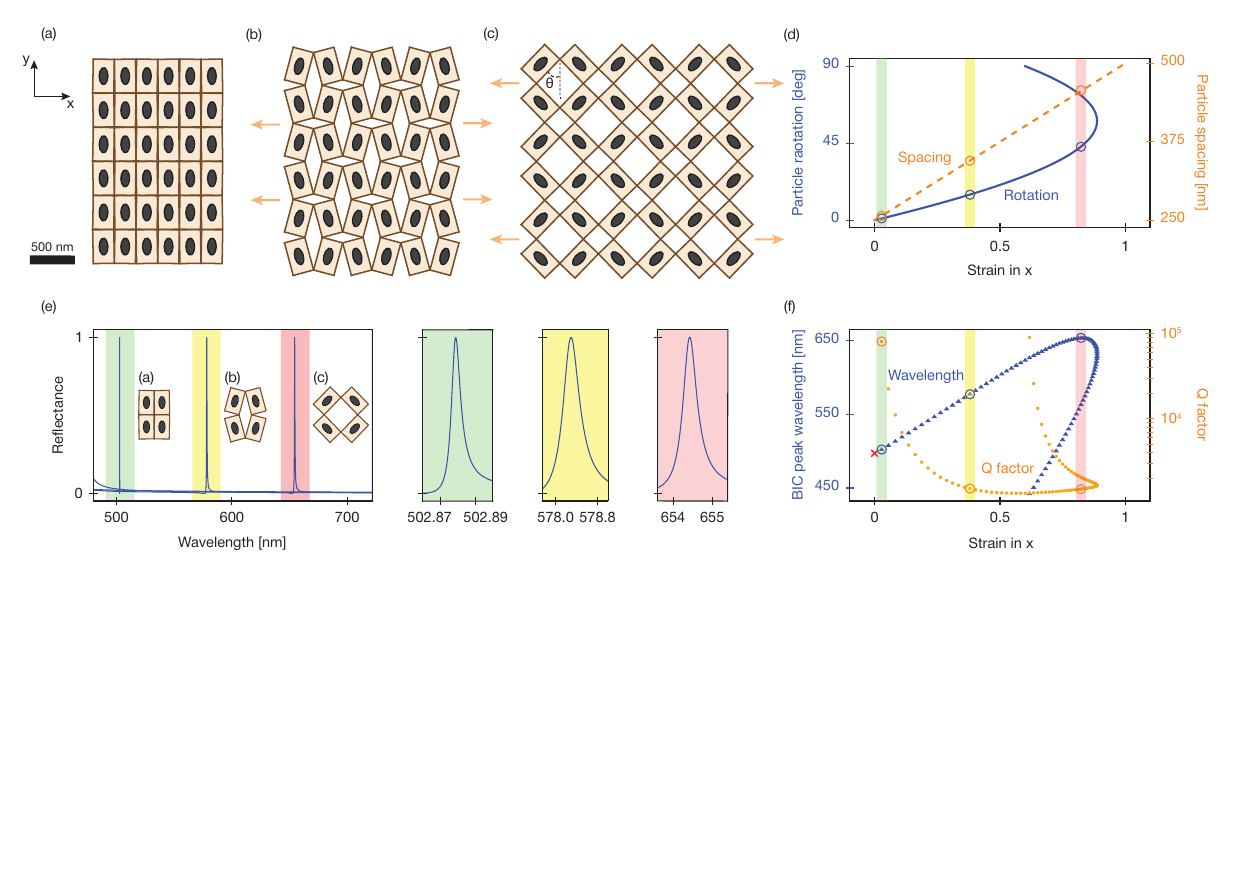}
\caption{\textbf{Mechanical deformation and tunable optical response of an idealized kirigami metasurface}. \textbf{(a-c)} Schematics of 3$\times$3 unit cell sections of the kirigami metasurface under increasing levels of strain: minimal strain ($\theta=1$\degree) (a), medium strain ($\theta=15$\degree) (b), and large strain ($\theta=44$\degree) (c). Each elliptical nanoparticle is centered on a rectangular tile measuring 250$\times$400 nm. \textbf{(d)} With increasing strain in the x-direction, the tiles \textemdash{} and thus the particles \textemdash{} are rotated by an angle $\theta$ (solid, blue) and displaced, leading to changes in the inter-particle spacing (dashed, orange). \textbf{(e)} Reflectance spectra for the arrays shown in (a-c) illustrating the giant tunability of the high quality factor resonant peak across the visible spectrum from green ($\sim$500 nm) through yellow ($\sim$580 nm) to red ($\sim$650 nm), with zoomed-in sections near these wavelengths to highlight the Fano lineshape. \textbf{(f)} Strain dependence of the resonance wavelength (blue triangles) and quality factor $Q$ (orange circles).}
\label{Fig1}
\end{figure*}

\emph{Idealized kirigami metasurface. ---}
We first introduce a reconfigurable metasurface made from resonant nanoparticles on a kirigami substrate that exhibits a tunable quasi-bound state in the continuum (q-BIC) with high quality factor ($Q$) by simple stretching (Fig. 1).  Unlike conventional stretchable materials such as polydimethylsiloxane (PDMS)~\cite{PDMSarticle}, judiciously placed cuts in the kirigami substrate enable engineered mechanical deformations and thereby unusually large strain and internal rotations (Fig.~\ref{Fig1}a-c) ~\cite{GRIMA,FlorijnPRL2014}. 

The kirigami is decorated by an array of nanophotonic resonators~\cite{firstellipses,Koshelev_BIC}---each elliptical resonator is placed at the center of each rectangular tile. Crucially, the periodicity of the metasurface and kirigami substrate must be commensurate. If we strain the kirigami along its x-axis, the angle $\theta$ (Fig.~\ref{Fig1}c) can vary from 0$\degree$  to $\sim$58$\degree$ at a strain of 0.89 (Fig.~\ref{Fig1}d), associated with an increase in the inter-particle distance up to 89\%. As such, a wide range of relative spacing and orientation between the resonators can be obtained on-demand by simply stretching the kirigami metasurface.  

\begin{figure}[b!]
%\hspace{-2.2cm}
\centering
\includegraphics[width=1\columnwidth,trim=0cm 0cm 0cm 0cm]{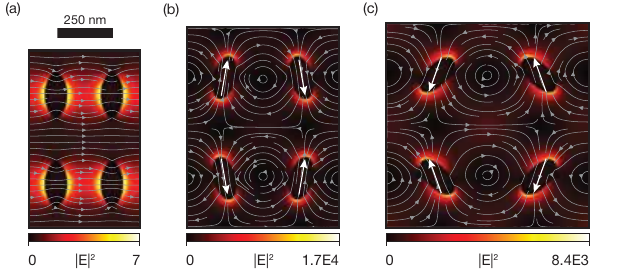}

\caption{\textbf{Electric field intensity profiles, normalized to the source intensity, for the idealized kirigami metasurface.} Gray (arrow) lines indicate the electric field lines in the plane. Overlaid white arrows indicate the effective dipole moments. \textbf{(a)} The undeformed state ($\theta=0$\degree) displays negligible field enhancement. \textbf{(b)} Small strain condition ($\theta = 15^\circ$) and \textbf{(c)} large strain condition ($\theta = 44^\circ )$ show enhanced localized field intensities.
}
\label{Fig2}
\end{figure}

\begin{figure*}[t!]
%\hspace{-2.2cm}
\centering
\includegraphics[width=2\columnwidth,trim=0cm 5cm 0cm 0cm]{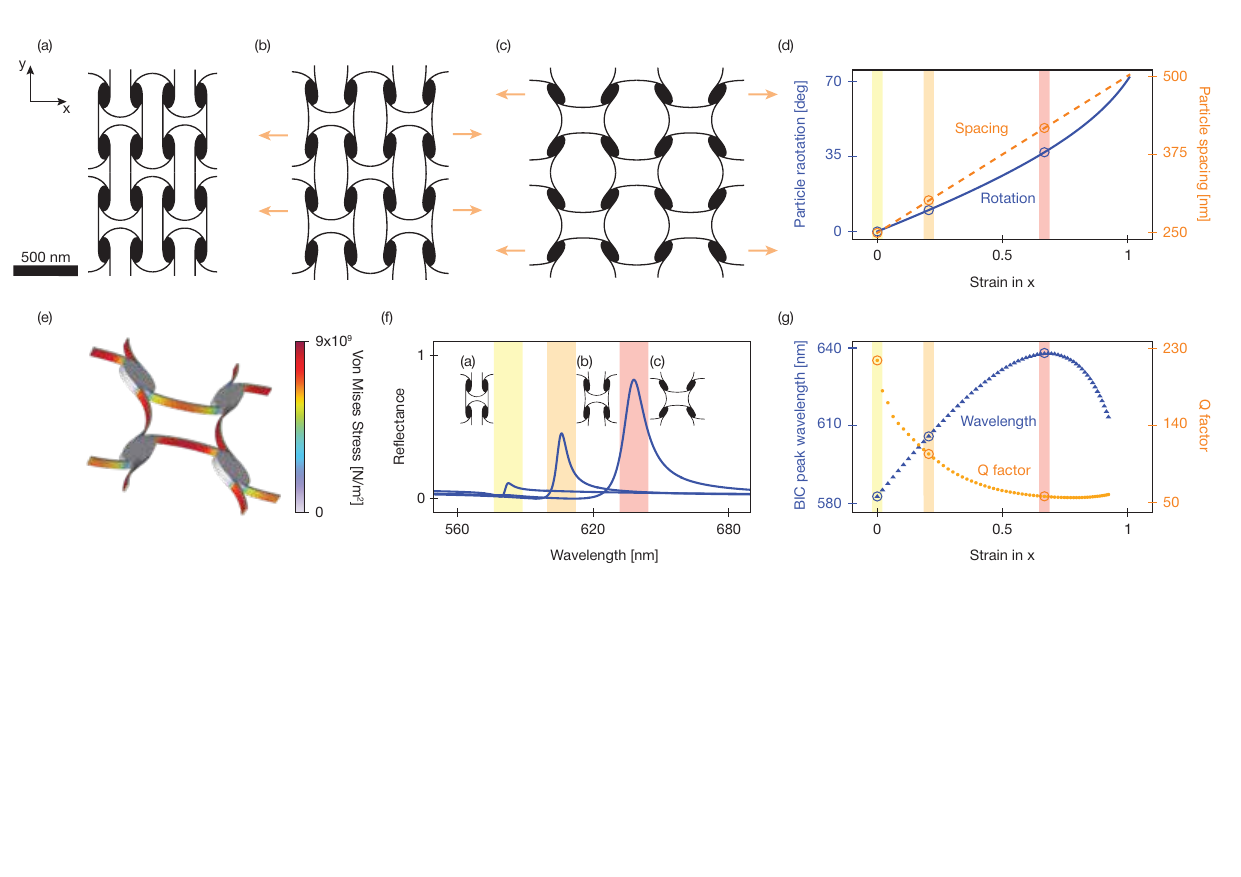}
\caption{\textbf{Mechanical deformation and tunable optical response of the beam-linked metasurface} \textbf{(a-c)} Schematics of 2$\times$2  unit cell sections of the kirigami-inspired metasurface: the initial undeformed state (a) shows no rotation of the elliptical nanoparticles, while moderate strain ($\theta=10$\degree) (b), and large strain ($\theta=38$\degree)  (c) show increasing rotation and inter-particle spacing.\textbf{(d)} Particle rotation angle $\theta$ (solid, blue) and displacement (dashed, orange) as functions of strain in the x-direction. \textbf{(e)} Simulated stress distribution for $\theta=60$\degree, showing strong localization of the stress at the beams. \textbf{(f)} Reflectance spectra for the arrays shown in (a-c), illustrating the large tunability of the resonant peak across the visible spectrum from yellow ($\sim$580 nm) through orange ($\sim$610 nm) to red ($\sim$640 nm). \textbf{(g)} Strain-dependence of the resonance wavelength (blue triangles) and quality factor $Q$ (orange circles), with a notable overall decrease in comparison to Fig.~\ref{Fig1}f.}
\label{Fig3}
\end{figure*}

\emph{Giant tunability. --- } To test whether stretching the kirigami leads to notable changes in the optical response, we use finite-element frequency domain simulations to calculate the reflection spectrum for varying metasurface geometries (see Methods in SI). The reflection spectrum of the kirigami metasurface (Fig.~\ref{Fig1}e) reveals a single resonant peak with three remarkable features. First, as the kirigami metasurface is stretched the spectral position is tuned across the spectrum from 500~nm to 650~nm (Fig.~\ref{Fig1}f) due to the strong dependence of the qBIC mode to the inter-particle spacing. This range corresponds to 328$\times$ the resonant linewidth. This giant tuning range is in stark contrast with the relatively small tunability of most existing platforms for active metasurfaces. 

Second, the resonance quality factor decreases from theoretically arbitrarily large for infinitesimally small deformation to a minimum of $Q=1334$ at $\theta = 24$\degree~(Fig.~\ref{Fig1}f). Interestingly, this decrease is less pronounced than the $Q \propto \sin^{-2} \theta$ that is characteristic for the perturbation of qBIC modal symmetries~\cite{Koshelev_BIC}. We attribute this to the changes in particle spacing associated with the particle rotation within the kirigami metasurface (See SI). Third, the maximum spectral shift is achieved for $\theta_{max}=44\degree$, which shows an unexpected discrepancy with the $58\degree$ that provides the largest displacement based on the kirigami design. 

\emph{Quasi-bound state in the continuum. --- } To elucidate the interplay between internal rotations and nonlocal resonances, we present in Fig.~\ref{Fig2} the electric field lines (arrows) and  intensity enhancement (color) at resonance for three levels of strain ($\varepsilon$; the unstretched metasurface ($\varepsilon=0$,  Fig.~\ref{Fig2}a), moderate strain $(\varepsilon=0.38$ Fig.~\ref{Fig2}b), and large strain $(\varepsilon=0.83$ Fig.~\ref{Fig2}c). The individual ellipsoidal particles support a strong electric dipolar Mie resonance, with the dipole moment (anti-)parallel to the major axis. However, for the unstretched situation the modal symmetry of the unit cell inhibits incident light from coupling to this pure BIC, and the dipolar field lines are not observable (Fig.~\ref{Fig2}a). The resulting field intensity in the plane of the metasurface is not significantly enhanced beyond the incident intensity: $|E|^{2} < 10 |E_{0}|^{2}$. For non-zero stretch ratios the unit cell expands, the particles rotate, and the modal symmetry of the unit cell is perturbed. For small angles (Fig.~\ref{Fig2}b), the electric field lines of the dipolar Mie modes are clearly observable in the individual particles, with strong field intensities concentrated at the particle tips. Careful evaluation of the field lines shows that within a four-particle unit cell neighboring dipole moments (white arrows guide the eye) are oriented anti-parallel with a relative angle close to $\theta$. Resonant coupling to this qBIC mode gives rise to strong field enhancements in plane of the metasurface ($|E|^{2} \ggg |E_{0}|^{2}$). Interestingly, for large strain, the effective dipole orientation within the individual resonant nanoparticles is no longer aligned with the long axis of the ellipsoidal particle, but is re-oriented towards smaller $\theta$ (Fig.~\ref{Fig2}c). This intuitively explains why the maximum spectral shift does not align with the maximum kirigami displacement. For $\theta = 90\degree$ the structure returns to a symmetry protected BIC where the dipoles align with the y-axis again (Fig.~S4), albeit at a lower resonance wavelength than the original BIC (Fig. \ref{Fig1}f).

\begin{figure}[b!]
%\hspace{-2.2cm}
\centering
\includegraphics[width=1\columnwidth,trim=0cm 0cm 0cm 0cm]{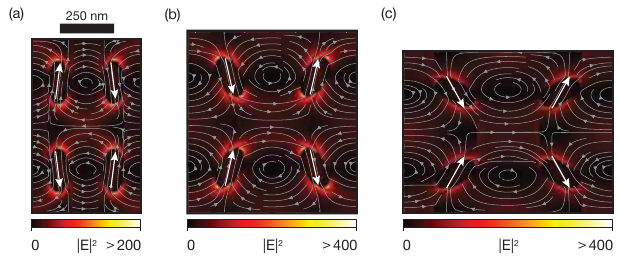}

\caption{\textbf{Electric field intensity ($|E|^2$) profiles, normalized to the source intensity, for the beam-linked metasurface.} Gray (arrow) lines indicate the electric field lines in the plane. Overlaid white arrows indicate the effective dipole moments. \textbf{a} The undeformed state ($\theta=0$\degree) already displays significant field enhancement. \textbf{(b)} Moderate strain condition ($\theta = 30^\circ$) and \textbf{(c)} high strain condition ($\theta = 60^\circ )$ also show field enhancement.
}
\label{Fig4}
\end{figure}

\begin{figure*}[t!]
%\hspace{-2.2cm}
\centering
\includegraphics[width=2\columnwidth,trim=0cm 0cm 0cm 0cm]{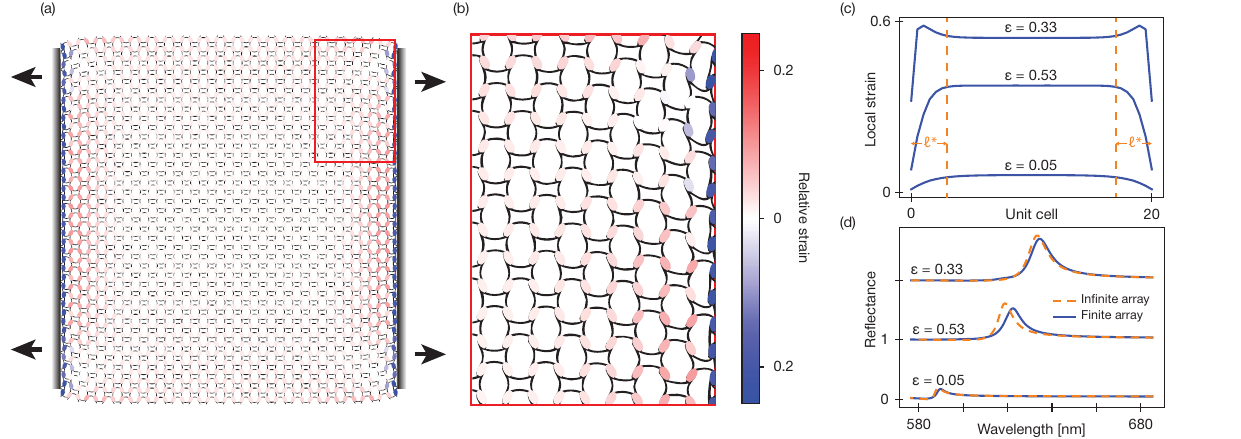}
\caption{\textbf{Finite size effects of the realistic kirigami metasurface}. (a) A finite size metasurface (20x20 unit cells) with rigid mechanical handholds on either side, showing variations in local deformation. The color of the particles indicates deviation in the local strain with respect to the bulk strain. (b) Zoomed-in view of the local strain distribution near the handholds highlighting localized deformation. (c) Local strain distributions across the structure for three strain values, each showing a strain plateau with roughly equal lateral size (between vertical dashed lines). (d) Comparison of the reflectance spectra between the finite array (solid blue curves) and an infinite array (dashed orange curves), illustrating how non-uniform stretching affects the overall optical properties, resulting in a reflectance peak with reduced amplitude and larger linewidth for the finite array.
}
\label{Fig5}
\end{figure*}

\emph{Beam-linked metasurface. ---} So far, we have explored the mechano-optical response of an ``ideal'' kirigami metasurface where the substrate was fictitious (i.e. $n=1$) and the refractive index of the silicon was constant and lossless. In the following, we aim to address these physical limitations in a second design: we include the dispersive and lossy optical constants of realistic silicon and the particles are mechanically supported. Crucially, our design is comprised of a nanopatterned 50~nm-thick silicon membrane that simultaneously functions as a mechanical metamaterial and as an optical metasurface (Fig.~\ref{Fig3}a-c). To this end, the optically-resonant nanoparticles are connected by carefully designed 10~nm wide beams. We optimize the shape of the beams such that: (i) they bend to emulate the counter-rotation mechanism of the rotating rectangle as closely as possible (see SI for a detailed discussion); (ii) the stress remains below the failure limit $\sigma_B< 20$GPa when the metasurface is stretched to a maximum of 60\degree (Fig.~\ref{Fig3}e). This is enabled by the notable enhancement of the strength of silicon at the nanoscale~\cite{tsuchiya2005evaluation}; (iii) they minimally affect the optical resonance of the nanoparticles.
Analogous to the ideal design (Fig.~\ref{Fig1}), applying in-plane strain initiates a pre-designed mechanical deformation of the particle's angle and inter-particle spacing (Fig.~\ref{Fig3}d). Just as for the idealized kirigami metasurface, the beam-linked metasurface's inter-particle spacing is directly proportional to the system strain. The particle rotation, however, is slightly increased over the relevant strain range in the beam-linked version. To assess the mechanical stability of the design, we perform simulations of the structural deformation and map the maximum stress throughout the unit cell (Fig.~\ref{Fig3}e). By virtue of the pre-curved connecting beams in the x-direction, the maximum stress is limited to $<10$~GPa for $\theta=60$\degree, confirming their mechanical stability under strain~\cite{tsuchiya2005evaluation}.

The introduction of optical damping and constrained mechanical deformations impact the optical response in four distinct manners: (i) The spectral tunability of the resonance is reduced to 580 -- 640~nm (Fig.~\ref{Fig3}f,g). Despite notably smaller than for the idealized kirigami metasurface, this spectral shift still corresponds to 4.7 times the largest resonance linewidth. (ii) The peak reflection is now limited by optical absorption and thus no longer reaches unity. (iii) The associated resonance quality factor is reduced from $Q=1336-\infty$ to $Q=55-215$ (Fig.~\ref{Fig3}g). Although this may seem as an undesired effect, a finite resonance bandwidth is a prerequisite for e.g. reflective displays that employ structural color to  provide sufficient optical power within the reflection peak. (iv) The introduction of the beams perturbs the unit cell symmetry and already enables weak coupling to the qBIC mode for $\theta=0$\degree - i.e. the mode is no longer bound (Fig.~\ref{Fig3}f).

To corroborate this, Fig.~\ref{Fig4} shows the field profiles on resonance for $\theta=0$\degree (a), $\theta=30$\degree (b), and $\theta=60$\degree (c). Indeed, despite the absence of particle rotation the field intensity is $>200$ the incident field as a result of formation of nanoscale hot spots at the beams. Also for non-zero $\theta$ the contact points of the beams with the particles form sharp corners that perturb the modal field profile and give rise to strongly localized hot spots due to the required continuity of the displacement field $D=\epsilon E$.

\emph{Finite size effects. ---} 
Boundary conditions are also known to play a a crucial role in mechanical metamaterials and in particular they are known to affect how internal rotations permeate into the bulk~\cite{Coulais_NatPhys2018}. Our beam-linked metasurface does not escape this rule and indeed we observe bulging at the edges, as a result of the non-slip boundary conditions (Fig. 5a,b). This leads to a non-homogeneous inter-particle distance throughout the material where the unit cells close to the edges don't stretch as much as the bulk of the metasurface (Fig. 5b). The spatial extent of this inhomogeneous deformation is further quantified by a characteristic length scale $\ell^*=1500$~nm (3 unit cells) (Fig. 5c), extracted from cross sections of the local strain throughout the 20 unit cells. The inhomogeneous mechanical deformation impacts the overall optical response of the metasurface. Averaged over the full metasurface, the reflection peak is slightly broader and exhibits a reduced amplitude (Fig. 5d). However, avoiding only the regions within a distance $\ell^*$ from the mechanical clamps will already retain the optical response of the bulk material (Fig. 5d). As such, the finite-size effects within the metasurface become negligible for all practical purposes.

\emph{Conclusion and outlook. ---} In summary,  we demonstrate how mechanical deformations in mechano-optical metasurfaces can tune a nonlocal qBIC across the visible spectrum. Our results pave the way for controlling light fields using designer mechanics, with the tantalizing prospect of more complex mechanical deformations and tunable optical functions for dynamic beam steering and wavefront manipulation beyond reflective filtering.

%\bibliography{references.bib} %Prints bibliography

\end{document}